\documentclass[aps,prb,twocolumn,superscriptaddress, showpacs]{revtex4}
\usepackage{graphicx} 

\begin{document}
 

\title{The nature of the magnetic and structural phase transitions in BaFe$_{2}$As$_{2}$}



\author{Stephen D. Wilson}
\affiliation{Materials Science Division, Lawrence Berkeley National Laboratory, Berkeley, California 94720}
\author{Z. Yamani}
\affiliation{Canadian Neutron Beam Centre, National Research Council, Chalk River Laboratories, Chalk River, Ontario, K0J 1P0, Canada}%
\author{C. R. Rotundu}%
\affiliation{Materials Science Division, Lawrence Berkeley National Laboratory, Berkeley, California 94720}
\author{B. Freelon}
\affiliation{Physics Department, University of California, Berkeley, California 94720}%
\author{E. Bourret-Courchesne}%
\affiliation{Materials Science Division, Lawrence Berkeley National Laboratory, Berkeley, California 94720}
\author{R. J. Birgeneau}
\affiliation{Physics Department, University of California, Berkeley, California 94720}%
\affiliation{Materials Science Division, Lawrence Berkeley National Laboratory, Berkeley, California 94720}

\date{\today}

\begin{abstract}
We present the results of an investigation of both the magnetic and structural phase transitions in a high quality single crystalline sample of the undoped, iron pnictide compound BaFe$_2$As$_2$.  Both phase transitions are characterized via neutron diffraction measurements which reveal simultaneous, continuous magnetic and structural orderings with no evidence of hysteresis, consistent with a single second order phase transition.  The onset of long-range antiferromagnetic order can be described by a simple power law dependence $\phi(T)^2\propto(1-\frac{T}{T_N})^{2\beta}$ with $\beta=0.103\pm0.018$; a value near the $\beta=0.125$ expected for a two-dimensional Ising system.  Biquadratic coupling between the structural
and magnetic order parameters is also inferred along with evidence of three-dimensional critical scattering in this system.   
\end{abstract}

\pacs{74.70.Dd; 75.25.+z; 75.50.Ee; 75.40.Cx}

\maketitle

\section{Introduction}
The recent discovery of high temperature superconductivity (high-T$_c$) within the iron pnictides \cite{kamihara} has resulted in a torrent of research into the electronic, magnetic, and structural behaviours of this new class of superconductors and their isostructural variants.  The pnictide superconductors now constitute a long sought comparator for testing the physics fundamental to the phenomenon of high T$_c$ superconductivity through comparison to the widely studied class of high T$_c$ lamellar copper oxides. The great deal of the interest surrounding this new iron pnictide class of superconductors originates from their striking parallels and notable differences relative to the high T$_c$ cuprates.  Chief among the similarities are the layered building blocks comprising the fundamental structure of these materials and the proximity of an antiferromagnetic ordered phase to the superconducting state within their respective phase diagrams\cite{cruz,zhaoce,birgeneaucupratereview}. However, it was also quickly uncovered that the parent compounds of the iron pnictides are poor metals, and studies of magnetism in these materials seem to indicate itinerant moments with significant three-dimensional (3D) character to the magnetic ordering\cite{mcqueeny,zhaoinelastic}---far from the two-dimensional (2D), localized moments of the $S=1/2$, antiferromagnetic Heisenberg behavior in the Mott insulating ground state of the cuprate parent systems\cite{birgeneaucupratereview}.  As has been pointed out recently, this does not preclude the importance of correlation effects in the iron pnictides whose electronic ground state may simply exist in close proximity to the Mott phase\cite{si, si2008}.  Whether the influence of electron-electron correlation effects within the underlying physics of the iron pnictides is as important as the known correlation effects in the cuprates stands as a central unresolved issue in the study of these new superconductors.   

Undoped, the parent compounds of the iron pnictides exhibit long range, three dimensional antiferromagnetic (3D AF) order with Fe$^{2+}$ moments aligned within the basal plane\cite{cruz, huang}.  The onset of AF order is preceded by (eg. LaFeAsO)\cite{cruz} or accompanied by (eg. BaFe$_2$As$_2$)\cite{huang} a structural phase transition from the high-temperature tetragonal structure to a low-temperature orthorhombic phase.  It has been argued that the structural phase transition is driven by the formation of long range magnetic order\cite{si,fang,barzykin}, and recent work has suggested a mechanism of fluctuating domain boundaries between energetically similar magnetic states to explain this effect even in systems where $T_s>T_N$\cite{mazinnaturephys}.  We will present an alternate model in this paper.  As magnetism is thought to play a role within the formation of the unconventional, superconducting ground state in these systems\cite{lee,cao,xu,han}, a comprehensive understanding of magnetic as well as the structural behavior in the iron pnictides, particularly in the undoped parent systems, is of considerable importance.     

The ThCr$_2$Si$_2$-type bilayer AFe$_2$As$_2$ (A=Ba, Ca, Sr, ...) systems, or 122-compounds, constitute one of the most promising avenues of experimental investigation of magnetism within the iron arsenides.  This is due to their relatively large Fe$^{2+}$ ordered-moments, and, most importantly, due to the fact that it appears to be straightforward to grow large single crystals of these materials.  All 122 variants studied to date exhibit a structural phase transformation from an $I4/mmm$ tetragonal to $Fmmm$ orthorhombic phase concomitant to the establishing of an AF phase with ordered moments ranging between $0.8\mu_{B}$ to $0.94\mu_{B}$ \cite{rotter,zhaoelastic,goldman,matan} per Fe-site---values rather smaller than those predicted from band-structure calculations\cite{yin,mazinprb,cao}.  Despite comparable moment sizes, there exists a significant range of $T_N$'s among these systems with $T_N=136$K in BaFe$_2$As$_2$\cite{matan}, $T_N=205$K in CaFe$_2$As$_2$\cite{goldman}, and $T_N=220$K in SrFe$_2$As$_2$\cite{zhaoelastic}.  This variation in $T_N$'s presumably arises from variations in the J$_{nn}$ exchange coupling between the different compounds.  Detailed studies of the magnetism within the parent phases of these 122-compounds, however, have yielded mixed results regarding the nature of both their structural and magnetic phase transitions.    

Neutron studies of the CaFe$_2$As$_2$ compound reported a first order magnetic phase transition at $T=173.5$K that was coupled to a simultaneous onset of the structural phase transition from tetragonal to orthorhombic symmetry\cite{goldman}.  Similar behavior was reported in SrFe$_2$As$_2$ at $T=205K$ where a first order magnetic phase transition coupled to the structural order parameter was observed \cite{zhaoelastic,jesche}. This claim was supported by heat capacity measurements which implied a discontinuous transition at the same temperatures\cite{krellner}, although separate heat capacity measurements of isostructural EuFe$_2$As$_2$ reported no hysteresis in the magnetic/structural phase transitions and speculated the possible presence of an undetectably small latent heat\cite{renEu122}.  However, a combined M\"ossbauer and x-ray diffraction study claimed to observe a continuous, second order structural phase transition in the same SrFe$_2$As$_2$ and EuFe$_2$As$_2$ systems \cite{tegel}. Neutron studies of BaFe$_2$As$_2$ polycrystalline samples initially reported a continuous onset of magnetic order below $T=143$K with a small hysteresis present in the accompanying structural phase transition\cite{huang}. This was followed by separate single crystal studies claiming a second order magnetic phase transition, albeit with substantially renormalized magnetic properties due to Sn incorporation within their crystal\cite{su}. Work by a different group on cleaner BaFe$_2$As$_2$ crystals failed to observe any hysteresis in the magnetic phase transition\cite{matan} in direct contradiction to separate NMR results showing a 4K hysteresis in the magnetic order parameter upon warming and cooling\cite{kitigawa}. A recent report has even claimed a hysteresis in excess of 20K through $T_N$, supporting a claim of a strongly first order phase transition in the same BaFe$_2$As$_2$ system\cite{kofu}.  Clear discrepancies amongst these various measurements mandate continued careful studies of the phase behavior in these 122-systems in an attempt to resolve the underlying nature of the magnetic ground state of the undoped, bilayer compounds. 

The fact that the magnetic phase transition in the BaFe$_2$As$_2$ system may be second order makes this compound an illuminative system for isolating the intrinsic nature of the phase behavior in this class of compounds.  Additionally, the substantially reduced $T_N$ in BaFe$_2$As$_2$ relative to its isostructural variants also makes it an interesting system for further study.  The reduced $T_N$ is potentially indicative of a stronger anisotropy in the spin behavior where 2D fluctuations within the FeAs layers renormalize the magnetic behavior.  For these reasons, we chose to study the spin and structural ordering within the BaFe$_2$As$_2$ compound via neutron diffraction.  

The remainder of this paper is organized as follows:  Section II discusses technical details of our sample preparation, neutron scattering experiments, and bulk transport measurements.  In Section III we present a preliminary characterization of our BaFe$_2$As$_2$ single crystals for relative crystal quality comparison with existing reports in the literature.  Section IV details our neutron diffraction measurements probing the magnetic and structural phase behaviors along with preliminary measurements of the critical scattering above $T_N$ in this system.  Section V presents our analysis of the critical behavior in this system and possible effects leading to discrepancies between various studies.  Finally, in Section VI we give our final conclusions and summarize our work.

\section{Experimental Methods}
For our experiments, we grew high quality single crystals of BaFe$_2$As$_2$ ( in orthorhombic notation $Fmmm; a=b=5.583\AA, c=12.947\AA$ at 250K) using a vertical Bridgman-type growth technique similar to that employed by Morinaga et al.\cite{morinaga}.  A 657 mg single crystal was obtained in a sealed quartz ampoule, 12 mm in diameter, in the presence of a FeAs flux. The complete details of the sample growth will be published elsewhere\cite{costelfuture}.  A 366mg single crystal, the larger half of a cleaved 657mg single crystal, was chosen for neutron studies.  Neutron experiments were carried out on the N5 spectrometer at the Canadian Neutron Beam Centre, National Research Council, Chalk River Laboratories.  The N5 spectrometer was used in a triple-axis configuration with a collimation of $30\acute{}-60\acute{}-$sample$-52\acute{}-72\acute{}$, unless otherwise indicated, and a fixed final energy of $E_f$=3.52 THz was used.  The $(0, 0, 2)$ reflections of pyrolytic graphite (PG) crystals were used for both the monochromator and the analyzer stages.   A pyrolytic graphite filter was used to reduce second order contamination from the incident beam.  

The BaFe$_2$As$_2$ crystal was mounted in the (H, 0, L)scattering plane (orthorhombic unit cell notation) within a closed-cycle refrigerator containing He exchange gas to ensure efficient heat exchange between the sample and the cryostat.  Neutron measurements were conducted between $3\leq T \leq 250$K in order to characterize the phase behavior.  In this work, positions in reciprocal space will be referenced using reduced lattice unit (r.l.u.) notation in which $Q(h, k, l)[\AA^{-1}]=(2\pi/a \times H, 2\pi/b \times K, 2\pi/c \times L) [r.l.u.]$.  Bulk magnetization and resistivity measurements were carried out using a Quantum Design physical property measurement system.  Our DC-magnetization measurements were performed on the smaller 291mg half of the cleaved crystal used in our neutron scattering studies while resistivity measurements were carried out on a small, $\approx10$mg, crystalline bar with dimensions 7x1.5x0.1 mm cut from a boule from a separate growth run.  

\begin{figure}
	\centering
		\includegraphics[scale=0.4]{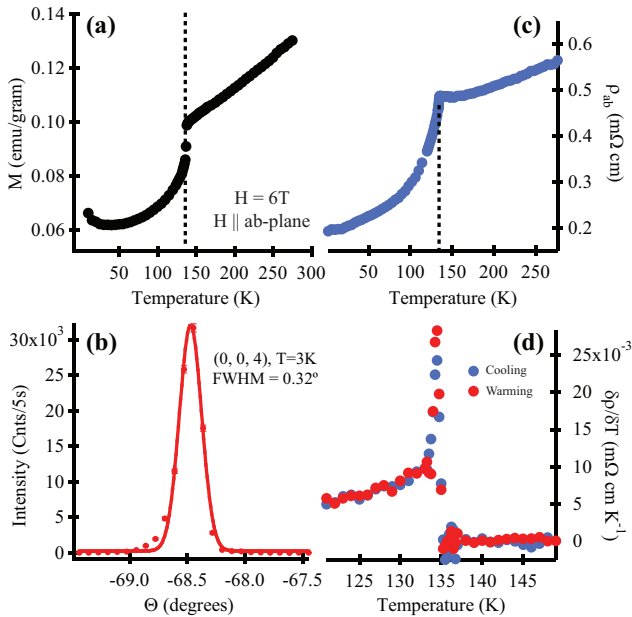}
		\caption{(a) Bulk susceptibility measurements showing the DC-magnetization measured with the magnetic field applied parallel to the $ab$-plane.  The dashed line denotes the AF transition temperature at $T=136$K.  (b) Neutron characterization showing a rocking curve illustrating the mosaic of our crystal.  The resolution limited peak full width at half maximum (FWHM) of 0.32$^\circ$ demonstrates the high quality of our single crystals.  The measurement was taken with collimation of $30\acute{}-12\acute{}-$sample$-12\acute{}-72\acute{}$. (c) Resistivity measurements on a small crystal of BaFe$_2$As$_2$ showing a linear decrease with decreasing temperature and a sharp drop at $T=135$K.  The drop in resistivity is indicative of the simultaneous magnetic and structural phase transitions in this system.  (d)  The derivative $\frac{\partial \rho}{\partial T}$ upon both warming and cooling showing a sharp peak indicative of the slope change at the transition temperature.  The peaks during both sample warming and cooling overlap perfectly showing no evidence for hysteresis in the simultaneous magnetic and structural phase transition of our samples. }
	\label{fig:Fig1}
\end{figure}

\section{Sample Characterization} 
Before presenting our neutron measurements, it is important to discuss first the quality of the BaFe$_2$As$_2$ samples studied in our experiments.  Magnetization measurements of BaFe$_2$As$_2$ often present the best gauge of sample quality; in the dc-susceptibility there emerge two salient features:  a Curie-type upturn in the susceptibility at low temperatures (typically below 30K), and a sharp decrease in the susceptibility curve at the antiferromagnetic ordering temperature.  A large susceptibility upturn in the low temperature Curie tail is reflective of the incorporation of magnetic impurities, often flux, into the sample volume. The inclusion of impurities into the sample also broadens and reduces the magnitude of the discontinuity in the susceptibility at $T_N$.  Our dc-susceptibility measurements of bulk samples grown via the Bridgman technique confirm this qualitative statement where progressive cuts starting at the larger diameter end of bullet shaped boules downward to the phase-pure single crystalline tip were taken.  These progressive cuts revealed a continual damping of the Curie-Weiss upturn at low temperature along with a sharpening of the inflection point near the magnetic phase transition as the single crystalline tip of the boule was neared.  Early work on BaFe$_2$As$_2$ crystals grown in Sn-flux, which is now known to incorporate Sn impurities into the BaFe$_2$As$_2$ matrix\cite{baek}, provides a dramatic demonstration of this effect where the dominant feature of the susceptibility is a continual Curie-Weiss-type upturn that persists until the lowest temperatures measured.  In typical magnetization measurements of these samples, any indication of magnetic ordering is typically obscured without subtraction of low-field data\cite{ni}.  

The dc-susceptibility of our BaFe$_2$As$_2$ crystal is shown in Fig. 1(a) where the susceptibility was measured with the field applied parallel to the $ab$-plane.  A linear decrease in the susceptibility from room temperature was observed until the AF ordering temperature of $\approx 137$K.  Below $137$K, there appears a sharp drop in the susceptibility followed by a minimal upturn in the susceptibility at temperatures below $\approx 25$K.  The sharp phase transition evident in our dc-susceptibility measurements combined with the excellent agreement between the magnetization curves of our bulk crystal and those of the smaller self-flux grown crystals in Ref. 34 demonstrates that our sample is of high quality and that the phase-pure section of the Bridgman boule had been reached.  In Section IV of this paper, we will report from our neutron measurements that the onset of magnetic order in this BaFe$_2$As$_2$ sample occurs at $136$K, and this $1$K difference in the magnetic onset temperatures is likely due to differences in the thermometry between the neutron diffractometer and magnetometer sample environments.

In order to characterize further our BaFe$_2$As$_2$ crystals, we also performed resistivity measurements in the temperature range from $T=2$K to $T=300$K.  The results of our measurements are shown in Fig. 1(c).  The resistivity, $\rho (T)$, continuously decreases upon cooling from $276$K with a sharp kink in the resistivity curve observed at the AF phase transition.  The residual resistivity ratio (RRR) gives $\rho (300K)/\rho (5K)=3.1$, a value consistent with earlier reports\cite{mandrus}.  The downturn in resistivity within this smaller sample occurs at approximately $T=135$K.  In Fig. 1(d) the first derivative of resistivity with respect to temperature, $\frac{\partial\rho}{\partial T}$, is shown, and it exhibits a sharp peak at $T=134.25$K as expected since, in a metal, $\frac{\partial\rho}{\partial T}$ scales like the heat capacity.  The temperature difference between the peak in the derivative of the resistivity and the downturn in the magnetic susceptibility is due to small variations in $T_N$ between different samples where subtle variations in the chemical gradient in different specimens can alter their antiferromagnetic ordering temperatures.  In order to perform preliminary tests for any hysteresis in the magnetic and structural phase transitions in this system, we measured the resistivity upon both slow warming and slow cooling the sample through $T_N$.  The results shown in Fig. 1(d) show no difference in the onset of the sharp slope-change in resistivity data upon warming and cooling with both peaks in $\delta\rho/\delta T$ coincident at $T=134.25$K.  This supports the notion that the magnetic phase transition in this material is, in fact, continuous. 

A further measure of our crystal's quality is provided by the mosaic of the crystal itself.  A rocking scan through the (0, 0, 4) reflection at 3K, shown in Fig. 1(b), exhibits a sharp peak with full width at half maximum (FWHM) of 0.32$^\circ$; a value determined by the resolution of the spectrometer.  This implies that the crystal quality is extremely high and that the crystal possesses a relatively small amount of strain-inducing defects and misalignments which may couple to and renormalize the intrinsic magnetic behavior.  This also allows for a more precise determination of the critical behavior in this system, where the absence of multiple grains within the mosaic avoids probing different crystallites with slightly differing magnetic properties.            

\section{Magnetic and Structural Phase Behaviors} 
\subsection{Magnetic order parameter} 
\begin{figure}
		\includegraphics[scale=0.4]{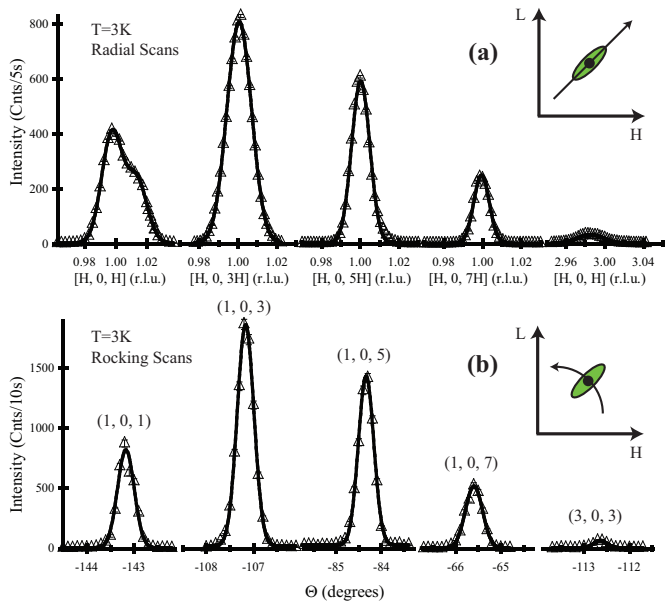}
		\caption{(a) Radial scans through magnetic peaks at the (1, 0, $L$); $L$=1,3,5,7 positions along with the (3, 0, 3) position.  (b) Rocking curves through the corresponding magnetic peaks in panel (a).  Insets in each panel qualitatively show the direction that the resolution ellipsoid intersects the magnetic Bragg position.}
	\label{fig:Fig2}
\end{figure}
 
\begin{figure}
		\includegraphics[scale=0.6]{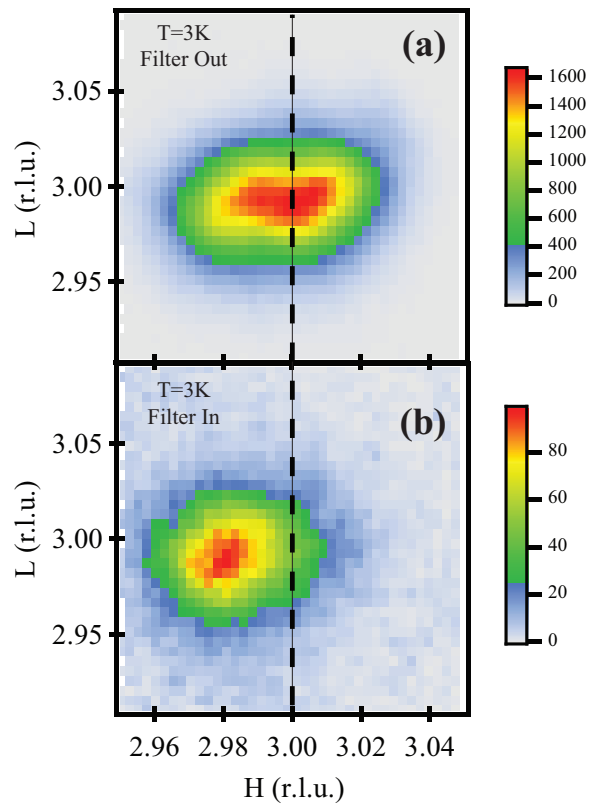}
		\caption{$\textbf{Q}$-mapping through mesh scans about the (3, 0, 3) magnetic Bragg reflection with (a) the PG filter out of the beam and (b) the PG filter in the beam.  The resulting nuclear scattering from the (6, 0, 6) position in panel (a) shows the structural twinning below $T_s$ which is absent in the purely magnetic scattering from the (3, 0, 3) reflection in panel (b).  This indicates that the moments are aligned along the longer $a$-axis in agreement with previous reports\cite{huang, zhaoelastic}}.
	\label{fig:Fig3}
\end{figure}

Turning now to our neutron diffraction results, Figure 2 shows scans at $T=3$K through the magnetic positions $\textbf{Q}=(1, 0, L); L=1,3,5,7$ and $\textbf{Q}=(3, 0, 3)$.  Panel (a) of Fig. 2 displays radial scans through these magnetic peaks, and panel (b) shows rocking curves through the same positions.  Insets in each panel qualitatively illustrate the direction that the resolution ellipsoid intersects the magnetic Bragg peaks within the corresponding scans.  The measured intensities are consistent with the published spin structure \cite{huang, zhaoelastic} for the 122 variants where the Fe-moments align within the basal plane.  The magnetic correlation length within the $ab$-plane is resolution limited with a minimum correlation length of $385\AA$ and a minimum out of plane correlation length of $350\AA$ determined with a collimation of $30\acute{}-12\acute{}-$sample$-12\acute{}-72\acute{}$.  In determining the direction that the moments lie within the $ab$-plane, the structural twinning which occurs below the tetragonal to orthorhombic phase transition was utilized as in Ref. 18.  Maps in $\textbf{Q}$-space were performed with both the PG filter in the beam and with the PG filter removed at the (3, 0, 3) position.  Figure 3 shows the results of these scans with the PG filter out (Fig. 3(a)) and the PG filter in (Fig. 3(b)).  In Fig. 3(a), the second order contamination of the beam dominates the scattering signal through $4E_i$ neutrons scattering from the (6, 0, 6) nuclear Bragg reflection.  The distended peak shape reflects the equal population of the twin (6, 0, 6) and (0, 6, 6) domains consistent with the expected twinning upon cooling into the orthorhombic phase.  Upon replacing the PG filter in Fig. 3(b), the second order contamination is strongly suppressed, and the scattering that remains originates from the (3, 0, 3) magnetic reflection.  Scattering from both twins no longer persists, and only one peak appears from the single twin aligned along the longer $a$-axis.  This verifies the $\vec{\textbf{Q}}_M$=(1, 0, 1) ordering vector for the spin structure in this system, consistent with previous reports \cite{kofu, su}.  Normalizing the (1, 0, 3) magnetic intensity to the (0, 0, 4) nuclear reflection yields an ordered moment of $0.93\pm0.06\mu_B$ per Fe$^{2+}$ ion.  In obtaining this normalization, the assumption was made that the structural twin domain populations are equal (an assumption supported by our subsequent analysis of the structural phase transition).

Having established that the low temperature magnetic structure is consistent with previous reports, we then examined the temperature dependence of the magnetic scattering at the strongest magnetic reflection at $\textbf{Q}=$(1, 0, 3).  Figure 4 shows the temperature dependence of the magnetic Bragg intensity upon both slowly warming and slowly cooling the sample. For both cooling and warming scans, a 20 minute equilibration was allowed between each temperature ramp of 1K with the cooling rate set to be 2K/min for the ramp itself.  The magnetic intensity at the peak (1, 0, 3) position was continuously measured during the equilibration time resulting in one measurement per $\approx50$mK.  This was followed by a full radial scan for each temperature step.  The temperature registered at the sample position was coupled to the crystal via helium exchange gas within the sample chamber of the displex rendering instantaneous measurements of the sample temperature accurate to within 10mK. In panels (a) and (b) within Fig. 4, the peak intensity measured at the (1, 0, 3) position is over plotted with the integrated area of radial scans through the (1, 0, 3) position as the sample is respectively cooled and heated through the magnetic phase transition.  The excellent agreement between the temperature evolution of the peak value measured at $\textbf{Q}=$(1, 0, 3) with the integrated radial peaks at discreet temperatures demonstrates that the measurements of the magnetic ordering at the peak position are not effected by any extrinsic effects such as temperature inhomogeneity within the sample and that there is no effect from a broadening or shift in peak position upon cooling/warming through the phase transition.  The temperature dependence of the $\textbf{Q}=$(1, 0, 3) peak intensity can therefore be used as a valid measure of the magnetic order parameter squared.  This allows for a higher temperature resolution near the phase transition.

We will consider first the integrated intensities of the (1, 0, 3) magnetic peak upon cooling and warming as shown in Fig. 4(c). There is no resolvable hysteresis and the onset of magnetic order is continuous.  This is illustrated clearly in Fig. 4(d) where the peak intensities upon cooling and warming are over plotted.  Fitting the temperature dependence of these peak intensities with a simple power law $\phi(T)^2\propto(1-\frac{T}{T_N})^{2\beta}$ gives an AF ordering temperature of $T_N=135.95\pm0.05$K upon cooling and $T_N=136.02\pm0.09$K upon warming.  As expected, they agree within the errors.  Given that there is no evidence for hysteresis in the magnetic ordering in our data, we then fit the combined warming and cooling data to same power law relation giving $T_N=135.98\pm0.04$K and $\beta=0.103\pm0.018$. This is plotted as a dotted line in Fig. 4(d).  This $\beta$ value is close to the value, $1/8$ expected for the universality class of the two dimensional Ising model\cite{brush} and is consistent with earlier reports by Matan et al. \cite{matan}.  Upon narrowing the range of temperatures fit above and below the transition, the fit $\beta$ value systematically increases toward the 2D Ising value. Reanalyzing $\phi(T)^2$ with a fixed $\beta=1/8$ renders the fit displayed as a solid line in Fig. 4(d) with a refined $T_N=136.23\pm0.025$K, a value within the error of the result from our power law fit using a free $\beta$ parameter.  Given the close statistical proximity of the convergence of the fit $\beta$ value to the 2D Ising exponent in the critical temperatures closest to the phase transition and the fact that the constrained 2D Ising fit diverges from the results of the refined $\beta=0.103$ fit only far from the transition temperature, it appears that the magnetic ordering is approximated reasonably well by the 2D Ising exponent.  It is also worth noting that the large range in reduced temperature over which the power law fit models the development of the magnetic order in Fig. 4(d) is often a feature of other low dimensional antiferromagnetic systems such as K$_2$NiF$_4$ \cite{birgeneauK2NiF4} (see Appendix for further discussion). 

To test the validity of the power law fit, we performed further fits probing a range of temperatures about $T_N$ and a range of reduced temperatures, $t$, below the phase transition.  First, we observed that, upon screening the temperatures closest to the phase transition and refitting the data, the extracted critical exponents were smaller than the $\beta=0.103$ rendered from the full fit of the data.  For example, fits considering only $T<130$K yield a $2\beta=0.171\pm0.012$ while fits considering $T<134$K generate a $2\beta=0.194\pm0.006$.  This implies that the smallness of the critical exponent is intrinsic and not due to a sharpening induced by an unseen first order jump in the order parameter close to $T_N$.  Additionally, we repeated the power law fits in a shortened range of reduced temperatures about the transition, and the results of these fits are plotted in Fig. 5 as solid yellow lines through the peak intensities of the (1, 0, 3) magnetic peak as a function of temperature.  Fits within $t\approx 0.10$ resulted in a $T_N=136.26\pm 0.18$K and $2\beta = 0.224\pm 0.015$, while fits within $t\approx 0.05$ result in a $T_N=136.16\pm 0.21$K and $2\beta = 0.212\pm 0.021$.  For both fits, only data below $T=135.5$K were included. These $\beta$ and $T_N$ values agree with the values rendered from the full fit to the data within the statistical errors and therefore demonstrate that our fit results are robust.  

In order to constrain the range over which we can preclude the existence of a weakly first order phase transition in this system, we performed a linear fit to the magnetic order parameter squared at the temperatures closest to the phase transition.  This is shown within the inset of the top panel in Fig. 5.  The break in the data between $T=135.8$K and $136.1$K simply constitutes a small temperature range where equilibrated data were not collected and not a discontinuity in the order parameter.   The linear fit in this region represents the linear trade-off which must occur between phases during the onset of a first order phase transition, and, from the fit plotted within the inset of Fig. 5, our data allow us to preclude such a linear trade-off below $135.5$K.  Therefore, any first order nature to the phase transition is intrinsically weak and buried within the small temperature interval $135.5\leq T\leq 136.0$K; however it would seem to have little consequence to the overall phase transition that appears to be fluctuation driven (as discussed further in Section IV C).            

\begin{figure}
	\includegraphics[scale=0.4]{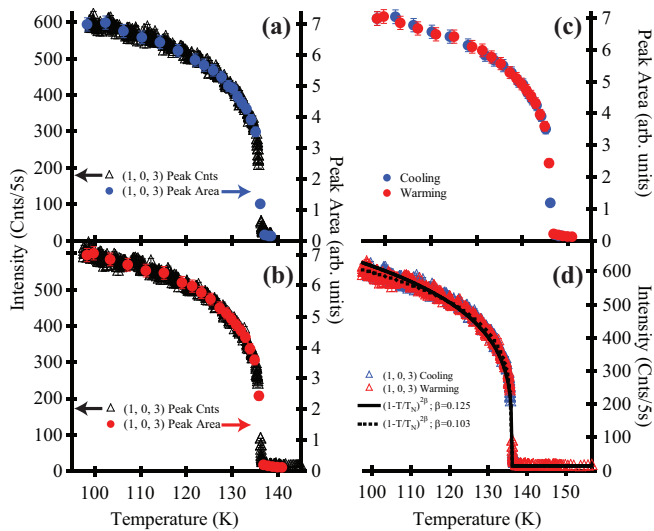}
	\caption{Measurements of the magnetic order parameter upon warming and cooling at the (1, 0, 3) reflection. Both the peak intensity measured at $\textbf{Q}=$(1, 0, 3) and the radially integrated (1, 0, 3) peak area are plotted upon both (a) cooling and (b) warming through $T_N$.  The perfect overlap between the peak intensity and integrated area measurements indicates that the peak intensity can be safely used as a measure of the magnetic order parameter, $\phi^2$.  (c) Integrated (1, 0, 3) peak areas overplotted as a function of cooling and warming.  There is no sign of hysteresis and the onset of magnetic order is seemingly continuous.  (d) Peak intensities at $\textbf{Q}=$(1, 0, 3) overplotted upon both cooling and warming the sample.  The dashed line represents the results of the power law fit described in the text with $\beta=0.103$ while the solid line displays the results of a power law fit with $\beta$ fixed at $1/8$.}
	\label{fig:Fig4}
\end{figure}

\begin{figure}
	\includegraphics[scale=0.55]{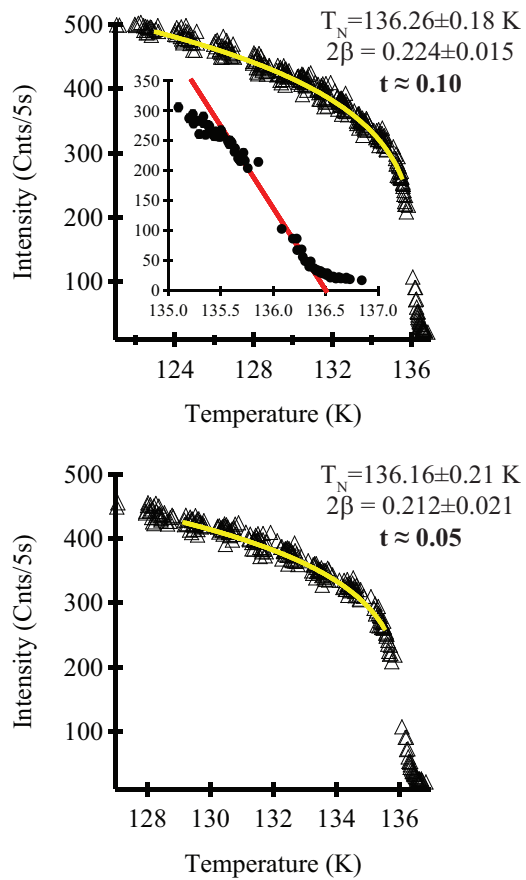}
	\caption{Continued analysis of the magnetic order parameter in different regimes of reduced temperature.  Top panel shows a power law fit to the $\textbf{Q}=$(1, 0, 3) peak intensity data in the reduced temperature regime $t\approx 0.10$ while the bottom panel shows a power law fit to the data in the $t\approx 0.05$ regime.  For both fits, only data below $T=135.5$K were included.  Power law fits are denoted by the solid yellow lines in both panels.  The inset in the upper panel replots the magnetic phase transition zoomed in to within $\pm 1$K of the phase transition.  A linear fit to the data between $135.6$K and $136.3$K is plotted as a solid line through the data and illustrates the maximum possible range for a linear trade-off between the paramagnetic and ordered magnetic phases required by a first order phase transition.}
	\label{fig:Fig5}
\end{figure}

\subsection{Structural phase transition}
Simultaneous to the onset of long range magnetic order, we observe a splitting of the (2, 0, 0) structural peak at $T_s=136.0$K indicative of the structural phase transition from $I4/mmm$ to $Fmmm$ symmetry and consistent with other observations of simultaneous magnetic and structural phase transitions in 122 variants \cite{huang, goldman, zhaoelastic}.  Due to resolution constraints and the intrinsically small structural distortion, the splitting of the (2, 0, 0) reflection could not be resolved into two distinct peaks; however, a clear broadening could be observed.  The broadened line shape may be effectively fit by assuming two resolution limited Gaussian peaks split symmetrically from the $\textbf{Q}=$(2, 0, 0) position.  This reflects the simply convolved projection of the resolution function along the $a$-axis with two intrinsically small nuclear peak widths approximated by delta functions.  The quality of the fits below and above $T_s$ is shown through the solid lines in Fig. 6(a), where the broadened peak at $T=98$K is fit by two Gaussian line shapes split by a parameter $\delta$ at (2$\pm\delta$, 0, 0) and at $T=140$K where the splitting is effectively $\delta=0$ .  The parameter $\delta(T)$ is directly representative of the structural order parameter via the relation  $\frac{\delta}{2}=\frac{a-b}{a+b}$. The structural order parameter squared ($\delta(T)^2$) is then plotted in Fig. 6(b) for both slow cooling and warming of the sample, using cooling/warming rates identical to those previously described in the measurement of the magnetic order parameter.  It is immediately apparent that there exists no observable hysteresis within the structural phase transition with both slow cooling and warming scans yielding an identical $T_s=136.0$K concomitant with the onset of magnetic order.

The smooth, non-hysteretic evolution of the structural order parameter through $T_s$ in Fig. 6(b) suggests a continuous transition coinciding with the onset of long-range antiferromagnetic order within this system and consistent with the theoretical understanding of a structural distortion coupled to the magnetic ordering\cite{mazinnaturephys, barzykin, si}.  To explore this further, we compared the temperature evolution of both $\phi(T)^2$ (the magnetic order parameter squared) and $\delta (T)^2$ upon both warming and cooling as shown in Figs. 6(c) and (d) respectively.  Both panels (c) and (d) display a remarkable agreement between the square of the magnetic and structural order parameters suggesting a strong biquadratic coupling term $\epsilon^2\phi^2$ between the two order parameters.  This contrasts with the behavior expected of the structural distortion as a secondary order parameter driven by the magnetic order which functions as the primary order parameter.  In that case there is an expected linear-quadratic coupling of the strain and staggered magnetization ($\propto\epsilon \phi^2$).  This makes the unexpected agreement between the $\delta(T)^2$ and $\phi(T)^2$ parameters particularly interesting, and the implications of this coupling will be discussed in Section IV.    

\begin{figure}
	\includegraphics[scale=0.35]{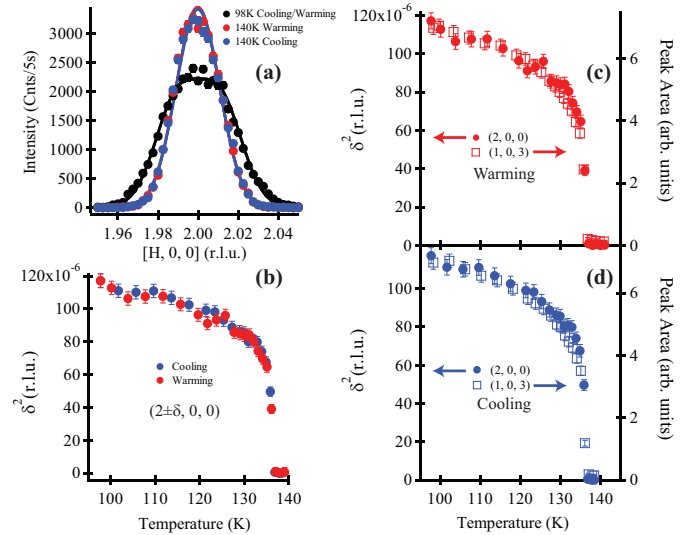}
	\caption{The evolution of the structural order parameter as a function of temperature.  (a) Longitudinal H-scans through the (2, 0, 0) nuclear Bragg reflection both below and above the structural phase transition $T_s$.  The solid lines are the results of fits of two Gaussians to the broadened peak below $T_s$ with the splitting, $\delta(T)$, determined as described in the text.  (b) $\delta(T)^2$ plotted upon both warming and cooling.  The onset of the structural splitting shows no hysteresis, identical to the development of the magnetic order in this material. $\delta(T)^2$ is overplotted with the integrated area of the (1, 0, 3) magnetic reflection as a function of (c) warming and (d) cooling showing a remarkable tracking between the structural order parameter squared and the magnetic order parameter squared.  The exact tracking within the errors verifies previous suggestions of a strong coupling between the onset of magnetic order and the structural lattice distortion in this material\cite{huang}}. 
	\label{fig:Fig6}
\end{figure}

\subsection{Critical scattering}
Given the apparently continuous nature of the magnetic order parameter within Figs. 4 and 6 and the lack of any observable hysteresis in the AF ordering temperature, we next searched for the presence of magnetic critical scattering near $T_N$.  The analyzer was left in for these measurements due to a gain in the signal to background ratio.  Hence any critical scattering observed does not represent an energy integrated response, but rather only the quasi-elastic critical scattering integrated over the spectrometer's energy resolution ($\Delta E=1.14$ meV FWHM). The results of our critical scattering measurements are shown in Fig. 7.  The scattering about the strongest magnetic reflection, $\textbf{Q}=$(1, 0, 3), was measured in the temperature range immediately above $T_N$ revealing a weak critical scattering signal extending up to $T\approx150$K (shown in the shaded area of Fig. 7(a)).  Each point in Fig. 7(a) represents the integrated area of radial scans through the (1, 0, 3) position, and select scans at temperatures above $T_N$ are plotted in Fig. 7(b).  The residual peak above $T=150$K results from remnant second order contamination in the incident neutron beam scattering from the nuclear (2, 0, 6) reflection.  This was determined by the lack of temperature dependence of the remnant (1, 0, 3) signal upon warming from $T=150$K to $T=250$K.

The dimensionality of the critical signal was measured via scans in Q-space through the (1, 0, 3) position probing spin correlations along the $a$-axis and correlations out of the $ab$-plane, along the $c$-axis.  The results of these $\textbf{Q}$-scans along $H$ and $L$ are plotted in Figs. 7(c) and (d).  In order to isolate the critical component of the peaks, the residual $\lambda/2$ nuclear peak was subtracted off using data collected at $T=148$K as the nuclear background.  Sharp, resolution-limited peaks centered at the $\textbf{Q}=$(1, 0, 3) position in both H and K scans above $T_N$ reveal the critical scattering to be fully three dimensional.  However, the widths of peaks in both $H$ and $L-$scans were resolution limited at all temperatures precluding any analysis of the magnetic correlation length as a function of temperature.  Additionally, scans at non-zero $\textbf{Q}_L$ offset from $L=3$ revealed no evidence for a substantial two dimensional component to the critical scattering.  This, however, does not preclude the existence of such two dimensional scattering where the signal may be too weak to be resolved due to the relatively small sample volume of our crystal.  Proper two-axis measurements integrating over a large range of energies on a larger sample or sample array would provide a more conclusive probe of any putative two dimensional critical scattering in this system.  

The existence of critical scattering above the phase transition supports the observation that the phase transition is second order; however, the strongly three dimensional nature of the critical scattering above $T_N$ contradicts the behavior expected for phase transitions within the 2D Ising regime.  Then, at the minimum, there must be a crossover from 2D to 3D critical behavior near $T_N$.  The three dimensional nature of the critical scattering can be reconciled with the anisotropic spin wave dispersion reported by Matan et al. \cite{matan} with a $c$-axis spin wave velocity of v$_c=57$meV$\AA$ relative to the in-plane spin wave velocity of v$_{ab}=280$meV$\AA$.  The two dimensional inelastic scattering above $T_N$ observed by Matan et al. \cite{matan} does not contradict our observation of three dimensional critical fluctuations driving the magnetic phase transition.  Indeed the observed behavior is similar to that of FeCl$_2$ \cite{birgeneaufecl2}, where the critical behavior was observed to be three dimensional despite a factor of 20 difference between in-plane and out-of-plane exchange couplings.

\begin{figure}
	\includegraphics[scale=0.4]{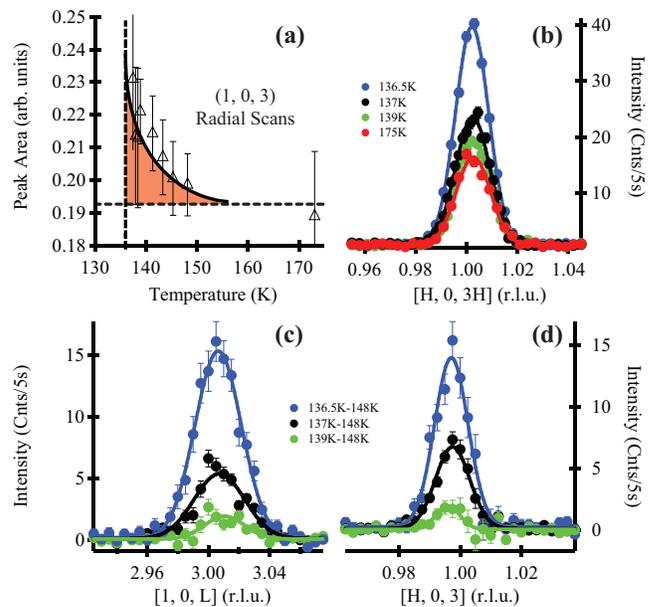}
	\caption{Critical scattering above $T_N$ at the (1, 0, 3) magnetic reflection.  (a) Radially integrated area of the (1, 0, 3) magnetic peak as a function of temperature.  The critical scattering above $T_N$ is denoted by the shaded area.  The dashed line denotes the level of $\lambda/2$ contamination in our measurements which can be treated as background. Radial scans at select temperatures are plotted in panel (b).  The large $\lambda/2$ peak apparent at 148K renders this measurement of critical scattering only preliminary.  $\textbf{Q}$-scans through (1, 0, 3) along both the $H$ and $L$ directions in reciprocal space are plotted in panels (c) and (d) respectively.  The nuclear background scattering from the 148K  $\lambda/2$ peak was subtracted off in order to isolate the magnetic portion of the peaks.  Sharp, resolution-limited peaks in both in plane and out of plane scans demonstrate that the critical scattering observed is fully three-dimensional.  The solid lines in panels (b)-(d) are the results of Gaussian fits to the data.}    
	\label{fig:Fig7}
\end{figure}

\section{Discussion}
The first issue to be addressed is the direct discrepancy between our neutron measurements revealing a seemingly second order magnetic phase transition in BaFe$_2$As$_2$ and those recently reported by Kofu et al.\cite{kofu} which claimed to observe a strong first order magnetic behavior.  One key factor which can potentially alter the magnetic phase transition in the iron pnictides is the coupling of the magnetic order to elastic strain within the system. It has long been known that through coupling to elastic strain or lattice deformation a second order magnetic phase transition can be driven first order\cite{bean}.  Until recently, this effect has not been taken into account in the iron pnictides; however, recent work by Barzykin and Gorkov\cite{barzykin} showed explicitly the ability of an external homogeneous deformation to alter dramatically the resulting magnetic phase behavior in the iron arsenides. The application of a lattice deforming strain can transform the order parameter through modifying the degree of Fermi surface nesting responsible for the formation of the SDW state, and, given sufficient strain, the Landau functional is modified to exhibit first order behavior.  However Barzykin and Gorkov \cite{barzykin} also predicted that the orthorhombic transition itself imposes an intrinsic quadratic magnetostriction term in the free energy which is alone sufficient to drive the transition to first order. Looking at the large variance of behavior between our measurements of magnetic ordering in BaFe$_2$As$_2$ and those performed by Kofu et al.\cite{kofu}, it is reasonable to infer that this intrinsic striction induced upon cooling through $T_S$ does not represent the dominant strain in this system. The samples in both our study and that by Kofu et al.\cite{kofu} should possess the same intrinsic magnetostriction terms upon cooling through the phase transition yet they yield vastly different outcomes.  This leads us to speculate that external strain imposed while mounting the crystal or sample dependent internal strain fields formed during growth due to imperfections or impurities plays the dominant role in modifying the phase transition.

Typical studies of 122 single crystals, consist of experiments on relatively thin, plate-like crystals with thicknesses on the order of 0.1-0.2 mm and much larger surfaces parallel to the $ab$-plane (often $\approx 5$mm$\times 5$ mm or greater\cite{goldman,zhaoelastic}).  This crystal geometry is especially susceptible to external strain upon mounting and possibly to the formation of internal strain fields due to dislocations during crystal nucleation\cite{fisherquantumoscillations}.  One notable effect of strain in these smaller samples is often manifested in the large asymmetry in their twin structural domain populations below $T_s$, where, by symmetry, there is no energetic difference between domains aligned along the orthorhombic $a$- and $b$-axes\cite{zhaoelastic}.  

Another factor which must be considered is the role of potential impurities within our BaFe$_2$As$_2$ sample. Impurities and imperfections have been observed to renormalize a first order phase transition into a continuous phase onset\cite{baek}.  One hallmark of this effect is a broadened critical regime remnant of the smeared intrinsic first order transition.  Figures 3 and 4, however, show a sharp transition in our crystal which reaches half of its saturation value below $T_N$ at $T_{50\%}<1$K.  Comparing this to previous studies of magnetic order in BaFe$_2$As$_2$ reveals that in fact our transition is much sharper than previous studies reporting discontinuous behavior where the $T_{50\%}\approx<20$K \cite{huang} in powder samples and $T_{50\%}\approx<7$K in single crystals\cite{kofu}.  This fact, combined with the sharp mosaic and clean transport properties of our sample, provides compelling evidence that the continuous transition we observe is not simply an impurity-induced renormalization of an otherwise first-order transition.  We therefore argue that strain or other extrinsic effects play the dominant role in renormalizing the phase transitions reported in previous studies of BaFe$_2$As$_2$ where the magnetic phase transition is broadened relative to the phase transitions observed in our work.  It is, however, not possible to preclude the existence of a weak first order transition within $0.5$K of $T_N$.  The presence of hysteresis in such a transition is difficult to detect given the small $\beta$ value of the power law behavior; however existing reports of hysteresis in excess of 4K\cite{kitigawa,kofu} would have been easily detected in our measurements.  

Additionally, a small first order hysteresis can be subtly smeared over a narrow temperature range in the critical regime generating a seemingly continuous response that is difficult to discriminate against.  As discussed in Section IV A, we performed additional fits in order to ensure that the small critical $\beta$ extracted from our measurements is not artificially low due to the influence of such a smeared first order transition through screening those temperatures closest to the magnetic phase onset.  In all cases, as the range of temperatures fit moved further away from $T_N$, the subsequently refined $\beta$ became progressively smaller than the $\beta=0.103$ extracted from the full fit of the data.  This strongly suggests that our observation of a sharp critical exponent consistent with the $\beta=1/8$ of the 2D Ising model is not an artifact from an unseen first order phase transition artificially sharpening the onset of the magnetic phase.      

A key result of our study is the unexpected behavior of the structural phase transition in BaFe$_2$As$_2$.  The simultaneous onset of both the structural and magnetic phase transitions in this system is consistent with the current experimental pictures of structural and magnetic order in the undoped 122 compounds; however, the increasingly prevalent speculation that the magnetic order drives the structural phase transition in BaFe$_2$As$_2$ is contradicted by our analysis of the relationship between the structural and magnetic order parameters.  Within the scenario of a secondary structural order parameter driven by the magnetic order through magnetoelastic coupling, the Landau free energy expansion for an order-disorder phase transition in an Ising system is of the form\cite{wolf}:  $F_{ME}= D_1\phi^2\epsilon + 0.5G_1\phi^2\epsilon^2+...$  The key term in this expansion is the linear-quadratic coupling term of the structural and magnetic order parameters respectively.  Given this term, the square root of the structural order parameter is expected to track the magnetic order parameter as a function of temperature.  Instead, as shown in Figs. 6(c) and (d), the structural order parameter tracks the magnetic order parameter directly.  This strongly suggests a biquadratic coupling term between the structural and magnetic order parameters in the free energy of this system and that the structural order parameter enters the free energy expansion as a primary order parameter, $\alpha(T)$.

The simplest free energy expansion of a system with coupled magnetic and structural order parameters is: $F_{tetra}= \phi^2(T)+\alpha(T)^2+2d\phi(T)^2\alpha(T)^2+\alpha^4(T)+\phi^4(T)+...$ \cite{liu,BirgeneauFeCoCl2} where $\phi(T)$ is the magnetic order parameter, $\alpha(T)$ is the structural order parameter, and $d$ is the coupling parameter between the two.  This free energy functional implies the existence of a tetracritical point in the phase diagram of the system. Tetracritical points have been previously observed within the phase diagrams of the high-T$_c$ cuprates through a biquadratic coupling of the spin density wave and the superconducting order parameters \cite{ylee}.  Theoretical treatment of this coupling generated exotic phase diagrams as a function of both doping and field tuning\cite{kivelsonstaged, demlerstaged}.  In the proximity of the tetracritical point, a mixed phase of coexistence between the two coupled phases $\alpha(T)$ and $\phi(T)$ exists below the intersection of their two second order phase transition lines, and the extent of this mixed phase within the phase diagram is a function of the coupling strength $d$.  The simultaneous onset of both the magnetic and structural phase transitions within BaFe$_2$As$_2$ strongly suggests that the system lies very close to the tetracritical point within its phase diagram.  While doping potentially changes many parameters simultaneously, the onset of the coupled phases should subsequently split as the system is tuned away from the tetracritical point regardless of the microscopic details of the system.  In fact, such a splitting has been recently reported in the phase diagram of Ba(Fe$_{1-x}$Co$_x$)As$_2$\cite{chupaper} which further supports the picture of a tetracritical point in this system.  Furthermore, this picture of two coupled primary order parameters in the close proximity of a tetracritical point supports the notion that the phase transitions are second order.  For an intrinsically first order transition in the transition region there must be a linear trade-off between one phase and another, and this could only happen in our case very close to T$_N$, which is not what is observed.           

The seeming contradiction between the $\beta=1/8$ two-dimensional Ising exponent modeling the critical regime of BaFe$_2$As$_2$ and the three dimensional critical fluctuations apparently driving the phase transition can be understood when the known asymmetry of the dimensional crossover within 2D Ising phase transitions is taken into account.  For a quasi-two dimensional Ising magnet with a finite coupling $R$ between the layers, there must exist a crossover to three dimensional magnetic behavior at temperatures close to phase transition.  The observation of three dimensional critical fluctuations above $T_N$ combined with the 2D Ising exponent modeling the development of magnetic order indicates such a crossover from two dimensional to three dimensional behavior in BaFe$_2$As$_2$.  If we assume that a measurable crossover to three dimensional phase behavior manifests itself at $t\approx 0.050$ in reduced temperature above $T_N$, then below $T_N$ such a crossover would persist to only $t\approx0.006$ in reduced temperature \cite{Jongh, Barouch}.  This is due to the known asymmetry in crossover regimes of the two dimensional Ising magnet close to the phase transition.   The small crossover temperature range below $T_N$ is insufficient to modify the overall power law fit to the critical behavior and the three dimensional behavior near the critical regime is effectively washed out by the much broader range of temperatures at which two dimensional behavior persists.  This conjecture would be confirmed by the observation of a crossover from 3D to 2D critical fluctuations above $T_N$.  

\section{Conclusion}   
In summary, we have studied both the magnetic and structural phase transitions within a large, high quality single crystal of BaFe$_2$As$_2$.  Our results reveal simultaneous, continuous magnetic and structural phase transitions at $T_N=136.0$K.  The lack of hysteresis in the order parameter onset temperatures and the detection of 3D critical scattering suggest that the phase transition is second order, although it is impossible to preclude a weak first order transition whose discontinuity is within $0.5$K of $T_N$.  Our measurements demonstrate that within bulk single crystal specimens of BaFe$_2$As$_2$ the substantially reduced strain from crystal mounting and improved crystal quality allows for the intrinsic phase behavior of BaFe$_2$As$_2$ to be resolved.  This is supported by consistent results between our bulk sample measurements and those by Matan et al. \cite{matan} whose sample was grown via similar methods.  We also report biquadratic coupling between the structural and magnetic order parameters which strongly suggests that BaFe$_2$As$_2$ is in the immediate vicinity of a tetracritical point.  Further work exploring the detailed doping evolution of the magnetic and structural phase transitions in this system is required to explore this scenario fully.  The critical behavior of the magnetic order parameter is well fit by the two-dimensional Ising critical exponent; however, the critical scattering immediately above $T_N$ is observed to be purely 3D.  Further experiments, with larger sample volumes, are required to make a definitive statement regarding the presence of any 2D component to the critical scattering as would be implied by the 2D Ising behavior observed below $T_N$.

\begin{acknowledgments}
We would like to thank A. Aharony and C. W. Garland for helpful communications.  This work was supported by the Director, Office of Science,  Office of Basic Energy Sciences, U.S. Department of Energy, under Contract No. DE-AC02-05CH11231 and Office of Basic Energy Sciences US DOE under Contract No. DE-AC03-76SF008.

\end{acknowledgments}

\end{document}